\documentclass[12pt]{article}
\usepackage{epsfig}
\usepackage{amsmath}
\begin{document}

 \title{Some Theoretical Estimations of Spatial Distribution of Compton Backscattered Laser
 Photons Beam. }
 \author{\sl Yu.~P.~Peresunko, \\
 National Science Center\\ Kharkov Institute of Physics and Technology,\\
 61108, Kharkov, Ukraine,\\
 E-mail: peresunk@kharkov.com
 }
 \date{}
 \maketitle

 \begin{abstract}
Spatial distribution of intensity, degree and direction of linear
polarization of tagged photon beam, which is obtained due to
Compton backscattering of laser light on high-energy electron
beam, are calculated. Effects of angular dispersion and spatial
spread of electron beam are taken into account. Calculations have
been carried out for the example of LEGS facility.
 \end{abstract}

 %\section{Theoretical description of the process of Compton scattering}
 Method of obtaining monochromatic and polarized photon beams of high energy
 and intensity by the Compton backscattering of laser photon on high-energy
 electron beam is widely used in many active and planned
 facilities.
 The most attractive feature of such photon beams is the
circumstance that obtained beam completely inherits polarizing
characteristics of an initial laser beam in a vicinity of the top
edge of a spectrum (Compton edge), and these characteristics can
be controlled well in this region. But in process of deviation
from Compton edge of spectrum the polarization of outgoing photons
decreases and shape of spatial distribution of intensity and
polarization becomes rather complicated. The photon beam could be
effectively split on two ones, linear polarization in the center
of photon spot is less than at its periphery.  This moment is
important for the analysis of experimental results, but it is
usually discussed at a level of scientific folklore, and the
purpose of the present note is - to discuss this effects in more
details.

Let's consider kinematics of the Compton scattering process
$$\gamma(k_1)+e^-(p_1)\to\gamma(k_2)+e^-(p_2)$$
 in the laboratory frame where axis $z$ coincides with a
beam's axis and $x$ axis belongs to horizontal plane of
accelerator's beam.  From 4--momentum conservation
 law $k_1+p_1=k_2+p_2$ it follows
 \begin{equation}
 \left( \alpha _x-\theta _x\right) ^2
 +\left( \alpha _y-\theta _y\right) ^2=r^2
 \label{eq.1}
 \end{equation}
 where

 $\alpha_x=\theta_1 \cos \varphi_1,\; \alpha_y=\theta_1 \sin \varphi_1$
 are plane angles in $x$ and $y$ directions of initial electron
 ($\theta_1,\,\varphi_1$ are its polar and azimuthal angles),
  $\theta_x=\theta_2 \cos \varphi_2,\; \theta_y=\theta_2 \sin \varphi_2$
 are plane angles in $x$ and $y$ directions of final photon
 ($\theta_2,\,\varphi_2$ are its polar and azimuthal angles).
  \[
 r^2=\gamma ^{-2}\left[ \lambda
 \left( \frac{\varepsilon _1-\omega _2}{\omega _2}
 \right) -1\right] ;\qquad
 \gamma =\frac{\varepsilon _1}{m};
 \qquad \lambda=2
 \frac{\left( k_1\cdot p_1\right) }{m^2}\approx \frac{ 4\omega_1
 \varepsilon_1}{m^2};
 \]
   $\varepsilon_1,\,m$ are energy and mass of initial electron,
 $\omega_1,\, \omega_2$ are energies of initial and final photons,
 correspondingly.
 Equation (\ref{eq.1}) is valid with an accuracy up to terms
\begin{equation}\label{eq.2}
   \left[ \frac{\omega _1}{\varepsilon _1},\quad \theta _2^2,
 \quad \theta _x^2,\quad
 \theta _y^2,\quad
 \gamma ^{-2}\right] \ll 1
\end{equation}
 Thus we can see from (\ref{eq.1}) that photons with a definite energy
 $\omega _2$
 are emitted at the surface of the circle cone with an axis along
 direction of initial electron motion
 and the opening angle
 $r$. Allowable energy of outgoing photon is determined from a condition
 of positive definiteness of $r^2$, and maximal possible energy
 of secondary photon is
 \begin{equation}
 \omega _{2\max }=\varepsilon _1\frac{\lambda}{1+\lambda};  \label{eq.3}
 \end{equation}
The cross section of the Compton scattering of the
 polarized photon by the unpolarized electron when one detects final photon
 with a Stoke's parameters
 $\xi_i^{(2)}$
 is \cite{Gr,peres}:
 \begin{equation}
 \frac{d^2\sigma }{d\omega_2d\varphi}=
 Sp\{ \widehat{\rho }_c\cdot \widehat{\rho }^{(2)} \}
 =\frac{r_o^2}{2 \varepsilon
 _1\lambda\left( 1+u^2\right) ^2\left( 1+\lambda+u^2\right) }\left\{\Phi_0+\xi _1^{(2)}
 \Phi_1+\xi _2^{(2) }\Phi_2+\xi _3^{(2)}\Phi_3\right\}   \label{eq.4}
 \end{equation}
 Here $\widehat{\rho }^{(2)}$ is a density matrix of detected final
 photon:
\begin{equation*}\label{ }
  \widehat{\rho }^{(2)}=\frac 12\left(
 \begin{tabular}{cc}
 $1+\xi_3^{(2) }$ & $\xi_1^{(2) }+i\, \xi _2^{(2) }$ \\
 $\xi _1^{(2) }-i\,\xi _2^{(2)
 } $ & $1-\xi _3^{(2) }$
 \end{tabular}
 \right) ;
\end{equation*}
and matrix $\ \widehat{\rho }_c$ can be considered as product of
cross section for unpolarized photon and proper density matrix of
final photon:
\begin{equation}\label{eq.5}
  \widehat{\rho }_c=\frac{r_o^2}{2 \varepsilon
 _1\lambda\left( 1+u^2\right) ^2\left( 1+\lambda+u^2\right) }\left(
\begin{tabular}{cc}
$\Phi_0+\Phi_3$ & $\Phi_1+i\,\Phi_2$ \\ $\Phi_1-i\,\Phi_2$ &
$\Phi_0-\Phi_3$
\end{tabular}
 \right) ;
\end{equation}
 where
 \begin{equation}
 \Phi _0 = 2+2\lambda+\lambda^2+u^2\left( 2+\lambda^2\right)
 +2u^4\left(1+\lambda\right) +2u^6
 -4\left( \xi _3^{(1)}\cos 2\varphi -\xi _1^{(1)}\sin 2\varphi
 \right) u^2\left( 1+\lambda+u^2\right)  \label{eq.6}
 \end{equation}
 \begin{equation}
 \Phi _1 = 2(1+\lambda+u^2)\left(\xi _1^{(1)}(-1+u^4\cos 4\varphi)+u^4\xi
 _3^{(1)}\sin 4\varphi
-2u^2\sin 2\varphi\right);  \label{eq.7}
 \end{equation}
 \begin{equation}
 \Phi _2 = -\xi _2^{(1)}\left( 1-u^2\right) \left(2+2\lambda+\lambda^2+2u^2(
 2+\lambda+u^2) \right);  \label{eq.8}
 \end{equation}
 \begin{equation}
 \Phi _3 = 2\left( 1+\lambda+u^2\right) \left(\xi _3^{(1)}(1+u^4\cos
 4\varphi)
 -\xi _1^{(1)}u^4\sin 4\varphi -2u^2\cos 2\varphi \right);  \label{eq.9}
 \end{equation}
In the expressions (4 -- 9) $r_0$ is classical radius of electron,
 $\varphi$ is azimuthal angle of outgoing photon,
which is counted from $x$ direction in the spherical reference
frame with polar axis along initial electron momentum, parameter
$\lambda=2 (k_1\cdot p_1)/m^2\approx 4 \omega_1 \varepsilon_1$,
and $u =\gamma r$.

Stoke's parameters  of initial laser photon $\xi^{(1)}_i$ and of
final photon $\xi^{(2)}_i$ are defined relatively to the axes
\{$x,y$\}, which are horizontal and vertical axes of electron
beam.
 According to general theory \cite{1,2} from (\ref{eq.4}-\ref{eq.5}) follows that
 proper Stoke's parameters of outgoing photon determined relatively to the
 laboratory frame axes $\left\{ x,y,z\right\} $ are:
 \begin{equation}
 \xi _1^{(f)}=\frac{\Phi _1}{\Phi _0},\quad \xi _2^{(f)}=\frac{\Phi _2}{\Phi
 _0},\quad  \xi _3^{(f)}=\frac{\Phi _3}{\Phi _0};  \label{eq.10}
 \end{equation}
 We shall use following parameterization of the photon's polarization
 characteristics \cite{3}:
 \begin{equation}
 \xi _1=P\cdot \cos 2\beta \cdot \sin 2\phi\, ;\qquad \xi _3=P\cdot \cos
 2\beta \cdot \cos 2\phi\, ;\qquad \xi _2=P\cdot \sin 2\beta  \label{eq.11}
 \end{equation}
 where $P=\sqrt{\xi _1^2+\xi _2^2+\xi _3^2}$-- is degree of the total
 polarization,

 $P_l=\sqrt{\xi _1^2+\xi _3^2}$-- is degree of the linear photon polarization,

 $\phi \ $ -- is the angle between axis $x$ and
 direction of maximal linear polarization of photon that is reading counter
 clockwise when one looks from the end of photon momentum vector.

 Let us consider what spatial distribution of tagged
photons with defined energy we shall obtain at the plane of target
situated at distance $L$ from Compton interaction region. We shall
suppose that (due to the large value of $L$ in comparison with
longitudinal size of laser and electron beams crossing region) the
Compton interaction points are distributed at one plane according
to Gaussian low with dispersions $\delta_x$ and $\delta_y$. The
plane angles $\theta_{1x}$, $\theta_{1y}$ of electrons relatively
to beam axes also have Gaussian distribution with dispersion
$\sigma_x$ and $\sigma_y$. So, spatial - angular density of
probability of initial electron states is:
\begin{equation}\label{eq.12}
  F(x_1,y_1,\theta_{1x},\theta_{1y})=\frac{1}{2\pi\sqrt{\delta_x
  \delta_y}}\exp{-\frac{x_1^2}{2 \delta_x^2}} \exp{-\frac{y_1^2}
  {2 \delta_y^2}}\;\frac{1}{2\pi\sqrt{\sigma_x
  \sigma_y}}\exp{-\frac{\theta_{1x}^2}{2 \sigma_x^2}}
   \exp{-\frac{\theta_{1y}^2}{2 \sigma_y^2}}
\end{equation}

 There is one-to-one correspondence between
 variables $(\omega_2,\;\varphi)$ and point $(x,\;y)$ of photon arriving at the
 target if set of variables $(x_1, y_1, \theta_{1x}, \theta_{1y} )$,
  which describe state of initial
 electron, are fixed:
\begin{equation}\label{eq.13}
  x = x_1+L \left(\theta_{1x}+r  \cos \varphi\right);\quad
   y = y_1+L\left( \theta_{1y}+ r \sin \varphi\right);
\end{equation}

The tagging system allows to select events with defined photon
energy, $\omega_2$. To take into account influence of tagging
system, we have to integrate differential cross section
(\ref{eq.4}) over photon energy with some appropriate distribution
which describes this system. Since accuracy of modern tagging
systems is rather high, near (1 - 2)\%, we can use as such
distribution a $\delta-$function. Then distribution at the target
plane of arrival points $\{x,y\}$ of tagged photons with energy
$\omega^0$, which have been produced at point $\{x_1,y_1\}$ by
Compton interaction laser photon and electron with plane angles
$\{\theta_{1x},\theta_{1y}\}$, is described by matrix:
\begin{equation}\label{eq.14}
  \widehat{\rho }_c^{(tag)}d\omega_2\,d\varphi= \widehat{\rho }_c
  \delta(\omega_2-\omega^0)\;d\omega_2\,d\varphi=
  \widehat{\rho }_c\delta(r^2-r_0^2)\frac{dx\,dy}{2L^2};
\end{equation}
where $ \widehat{\rho }_c$ is defined in (\ref{eq.5}-\ref{eq.9}),
 \begin{eqnarray*}
&&r^2 = \left( \frac{x-x_1}{L}-\theta_{1x}\right)^2+\left(
\frac{y-y_1}{L}-\theta_{1y}\right)^2,\quad
 \cos \varphi=\left( \frac{x-x_1}{L}-\theta_{1x}\right)/r; \nonumber \\
&&  r_0^2=\gamma ^{-2}\left[ \lambda
 \left( \frac{\varepsilon _1-\omega _2}{\omega _2}
 \right) -1\right]_{\omega_2=\omega^0};
 \end{eqnarray*}
To obtain desired final distribution we have to integrate
distribution (\ref{eq.14}) over states of initial electron with
distribution (\ref{eq.12}). In result we have:
\begin{equation}\label{eq.15}
  \langle  \widehat{\rho }_c^{(tag)}\rangle dx dy = \frac{dx dy}{2 \pi
  \sqrt{\Delta_x \Delta_y}}\int_0^{2 \pi}d\varphi\;
  \exp{-\frac{(x-r_0 L \cos \varphi)^2}{2 \Delta_x^2}}
  \exp{-\frac{(y-r_0 L \sin \varphi)^2}{2 \Delta_y^2}}\; \widehat{\rho
  }_c
\end{equation}
where $\widehat{\rho}_c$ is defined in (\ref{eq.5}--\ref{eq.9}),
and
\[
\Delta_x = \sqrt{(L \sigma_x)^2+\delta_x^2};\quad \Delta_y =
\sqrt{(L \sigma_y)^2+\delta_y^2};
\]

In other words, contribution to events, when at the point
$\{x,y\}$ comes photon with defined energy $\omega_2$, can give
whole set of initial electron directions which belongs to circle
cone with opening angle $r_0$. Therefore, to obtain polarization
characteristics of photon beam  we have to use integrated matrix
$\langle  \widehat{\rho }_c^{(tag)}\rangle$. Thus, we have for
distribution $I(x,y)$ of final photons with energy $\omega_2$
intensity following value:
\begin{equation}\label{eq.16}
  I(x,y)=  \frac{J \,r_o^2 \;\langle\Phi_0\rangle}{2 \varepsilon
 _1\lambda\left( 1+u^2\right) ^2\left( 1+\lambda+u^2\right) };
\end{equation}
where $J$ is appropriate current which describes intensity of
electron and laser beams.

For distribution of degree of final photons linear polarization
$P(x,y)$ we have:
\begin{equation}\label{eq.17}
  P(x,y) = \sqrt{\left(\frac{\langle\Phi_1\rangle}{\langle\Phi_0\rangle}\right)^2 +
  \left(\frac{\langle\Phi_3\rangle}{\langle\Phi_0\rangle}\right)^2}:
\end{equation}
and distribution of angles $\phi(x,y)$ between direction of linear
polarization and axis $x$ can be defined from conditions:
\begin{equation}\label{eq.18}
\langle\Phi_1\rangle/\langle\Phi_0\rangle = P(x,y) \sin{2
\phi(x,y)};\quad \langle\Phi_3\rangle/\langle\Phi_0\rangle =
P(x,y) \cos{2 \phi(x,y)};
\end{equation}
Operation $\langle...\rangle$ in (\ref{eq.16} - \ref{eq.18}) means
integration (\ref{eq.15}).

 At the figures (1 - 13) results of calculations on the
base of formulae (\ref{eq.16} - \ref{eq.18}) are depicted. This
calculations have been carried out for example of LEGS facility.
In this case energy of electron beam $\varepsilon_1 = 2580$ MeV,
UV laser photons have energy $\omega_1 = 3.7$ eV, so $\lambda = 4
\varepsilon_1\omega_1 = 0.146$ and upper limit of backscattered
photons' energy is $\omega_{2 max} = 330$ MeV. Dispersion of
electron beam in horizontal direction is $\sigma_x = 240\,\mu$rad,
and in vertical direction -- $\sigma_y = 80\,\mu$rad. Dispersion
of $\{x_1, y_1\}$ distribution of Compton interaction points is
$\delta_x = \delta_y =0.1$ cm.The distance  from Compton
interaction region to the target is $L=45$ m. It is supposed that
laser photon beam has 100\% linear polarization in horizontal
plane ($\xi_3^1=+1$) or in vertical plane ($\xi_3^1=-1$).
 \begin{figure}[ht]
\begin{center}
\includegraphics[width=0.48\textwidth]
{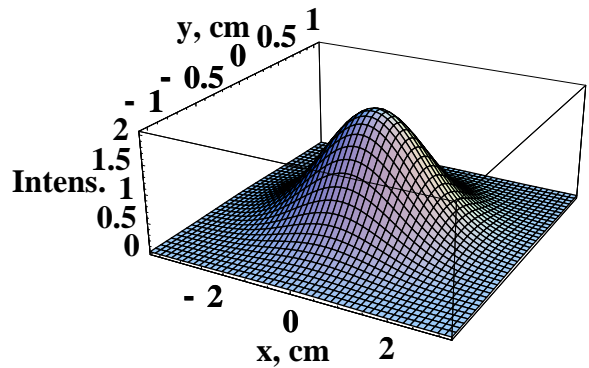}
 \hfill
 \parbox[t]{0.75\textwidth}{\caption {Spatial distribution of
 photon beam intensity. (Arbitrary units.) Photon energy
 $\omega_2 = 330$ MeV, $\xi^1_3 = +1$.
} \label{fig1}}
\end{center}
\end{figure}

At the figure (1) a spatial distribution of intensity of photons
near upper edge of spectrum is shown. One can see that this
distribution practically repeats Gaussian distribution of initial
electron beam. At the figures (2 -- 3) distributions of linear
polarization degree for photon energy $\omega_2 = 320$ MeV are
shown. Figure (2) corresponds to the case when initial laser
photons are polarized in horizontal plane ($\xi^1_3 = +!$), and
figure (3) --- to the case when laser photons are polarized in
vertical plane ($\xi^1_3 = -!$). One can see that polarization of
final photons is near 98.9\% for this energy but in the first case
polarization in the center of photon beam spot is slightly less.

The essentially other situation arises when photon energy
decreases below 280 MeV. Then in the case of horizontal
polarization of laser ($\xi^1_3 = +1$) secondary photon beam is
split into two ones and degree of polarization has deep minimum at
line $y=0.$ (See figures 4 - 5).

After change of laser polarization direction this splitting of
secondary beam disappeared and distribution of polarization degree
has a saddle shape. (See figures 6 - 7). Let's note that these
effects have been observed and reported in \cite{4}.
\begin{figure}[t]
\includegraphics[width=0.48\textwidth]
{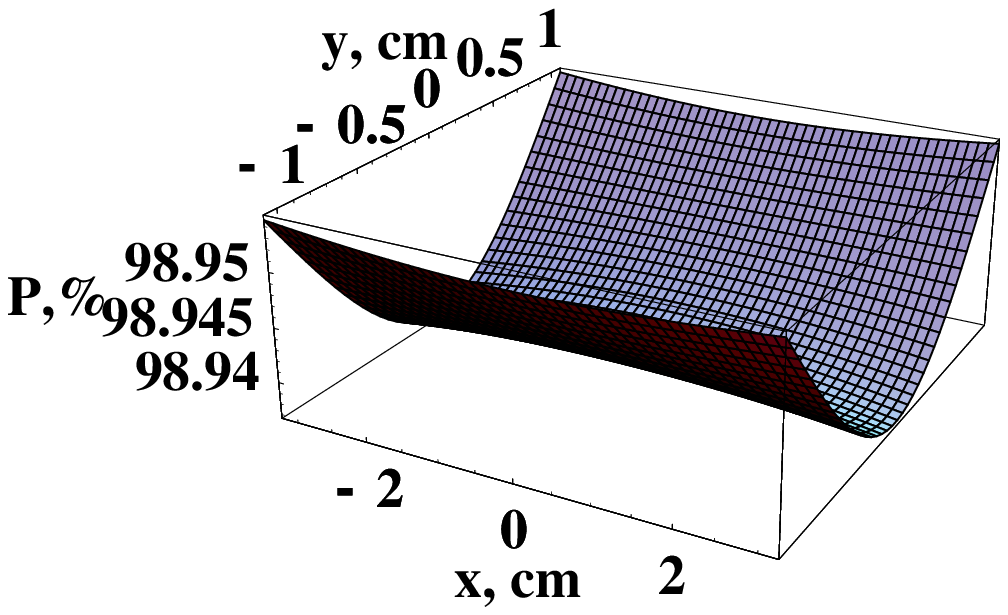}
 \hfill
\includegraphics[width=0.48\textwidth]
{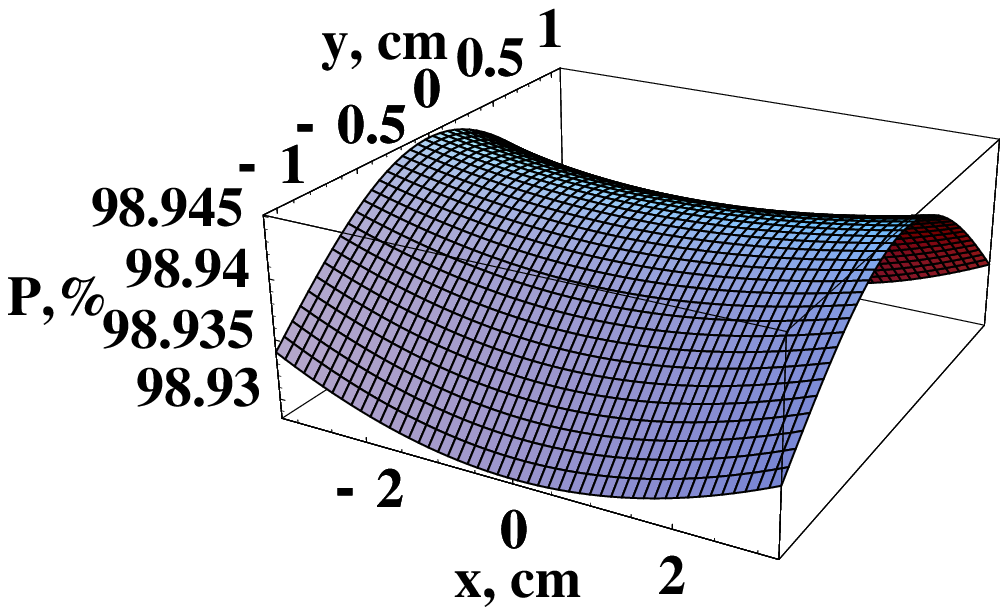}
 \hfill
 \parbox[t]{0.48\textwidth}{\caption {
 Degree of photons' linear polarization, $P$, \%. Photon energy
 $\omega_2 = 320$ MeV, $\xi^1_3 = +1$.
} \label{fig2}}
 \hfill
\parbox[t]{0.48\textwidth}{\caption{
Degree of photon's linear polarization, $P$, \%. Photon energy
 $\omega_2 = 320$ MeV, $\xi^1_3 = -1$.}\label{fig3}}
\end{figure}
 \begin{figure}[ht]
\includegraphics[width=0.48\textwidth]
{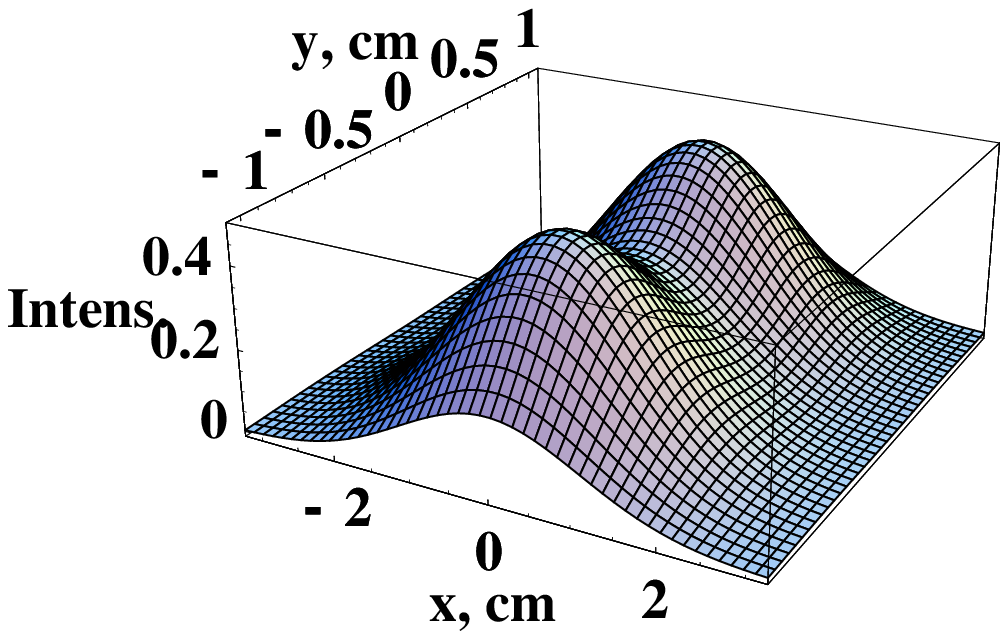}
 \hfill
\includegraphics[width=0.48\textwidth]
{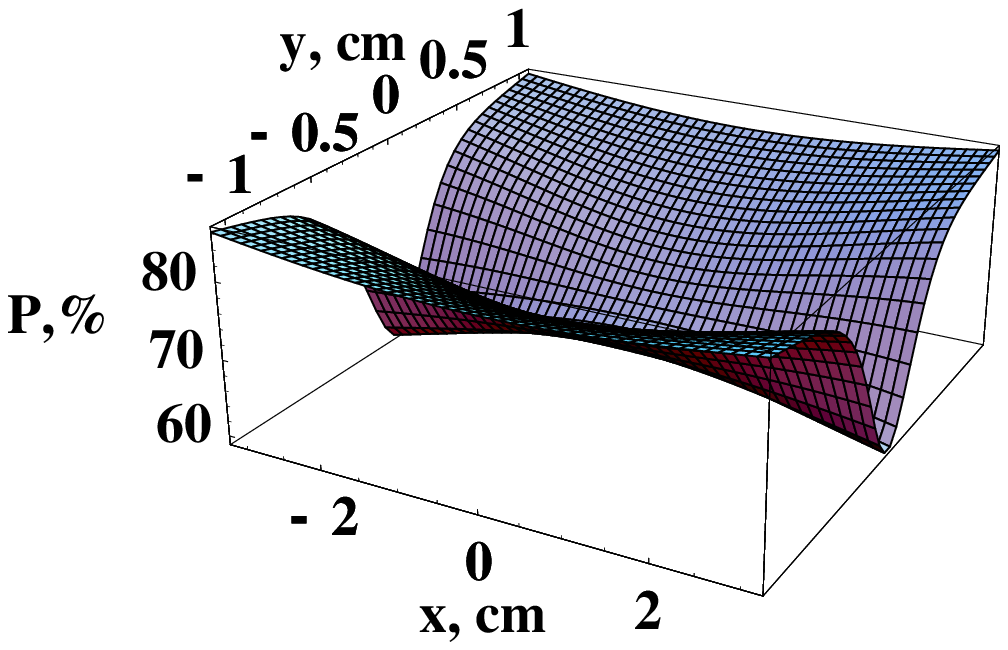}
 \hfill
 \parbox[t]{0.48\textwidth}{\caption {Spatial distribution of
 photon beam intensity. (Arbitrary units.) Photon energy
 $\omega_2 = 220$ MeV, $\xi^1_3 =+1$.
} \label{fig4}}
 \hfill
\parbox[t]{0.48\textwidth}{\caption{
Degree of photon's linear polarization, $P$, \%. Photon energy
 $\omega_2 = 220$ MeV, $\xi^1_3 = +1$.
 }\label{fig5}}
\end{figure}
 \begin{figure}[ht]
\includegraphics[width=0.48\textwidth]
{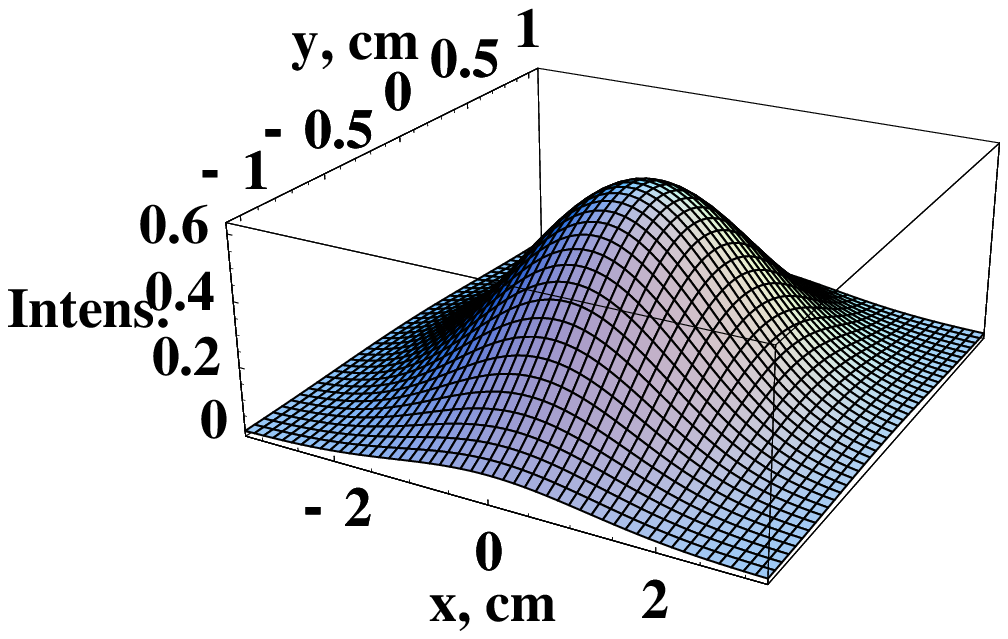}
 \hfill
\includegraphics[width=0.48\textwidth]
{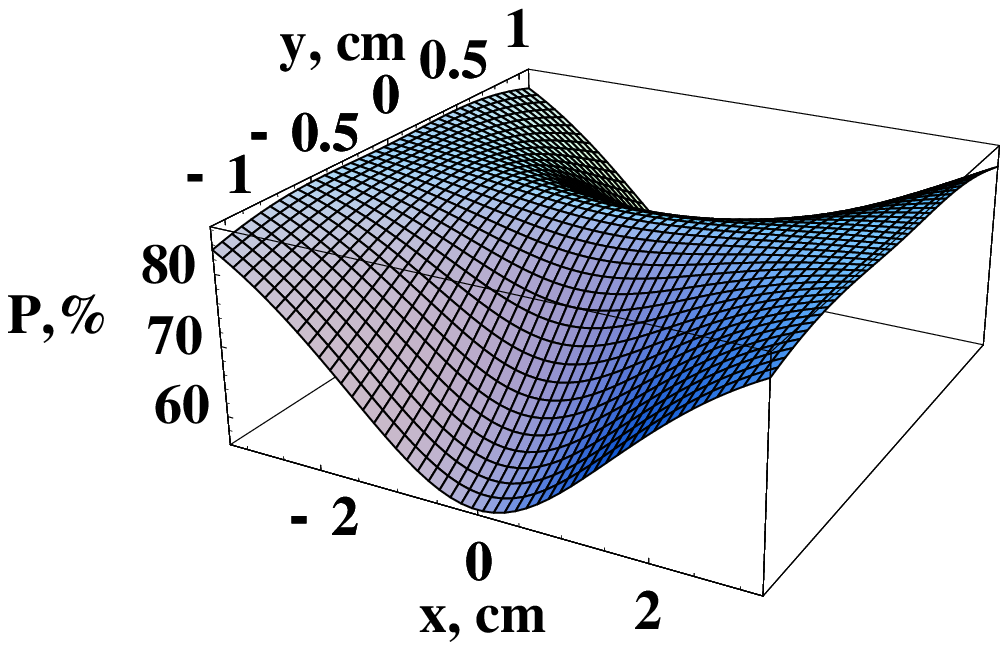}
 \hfill
 \parbox[t]{0.48\textwidth}{\caption {Spatial distribution of
 photon beam intensity. (Arbitrary units.) Photon energy
 $\omega_2 = 220$ MeV, $\xi^1_3 = -1$.
} \label{fig6}}
 \hfill
\parbox[t]{0.48\textwidth}{\caption{
 Degree of photons' linear polarization, $P$, \%. Photon energy
 $\omega_2 = 220$ MeV, $\xi^1_3 = -1$.
 }\label{fig7}}
\end{figure}

In some types of experiment with the purpose of statistics
increasing the events of secondary photons in some interval of
photon energy $\omega_2^{(2)}>\omega_2>\omega_2^{(1)}$ are taken
into account. In this case to obtain resulting intensity and
polarization we must integrate matrix $\widehat{\rho }_c^{(tag)}$
not only over spatial and angular distribution of initial
electrons but also over appropriate photon energies interval. Then
we shall have for spatial distributions of intensity $I(x,y)$,
degree of polarization $P(x,y)$ and for angle of polarization
direction $\phi(x,y)$ the same formulae as (\ref{eq.16} --
\ref{eq.18}) but averaging operation $\langle...\rangle$ have to
be replaced into $\langle...\rangle_\omega$ where
 $$
 \langle \widehat{\rho
}_c^{(tag)}\rangle_\omega=\int_{\omega_2^{(1)}}^{\omega_2^{(2)}} d
\omega \langle  \widehat{\rho }_c^{(tag)}\rangle;
 $$
  At the figures (8 - 9) such summarized distribution of intensity
  and degree of linear polarization for the case of horizontal
  polarization of laser photons is shown. At the figures (10 - 11)
  analogous distribution for the case of vertical polarization of
  laser photons are shown.

  At last, at the figures (12 - 13) the averaged over interval
   of photon energies 220 MeV$<\omega_2<$330 MeV
   spatial distribution of the
  angle between axis $x$ and direction of linear polarization of final
  photons $\phi(x,y)$ are shown for the cases of horizontal (fig. 12) and
  vertical (fig.13) polarization of initial laser photons.

  From these figures one can see that $\phi(x,y) = 0$ at the lines
  $x=0$ and $y=0$, but at periphery of photon spot it can reach $5^\circ$
  relatively to direction of initial laser polarization
   in the case of horizontal and $10^\circ$ in the case of
   vertical polarization of laser.
  \begin{figure}[h]
\includegraphics[width=0.48\textwidth]
{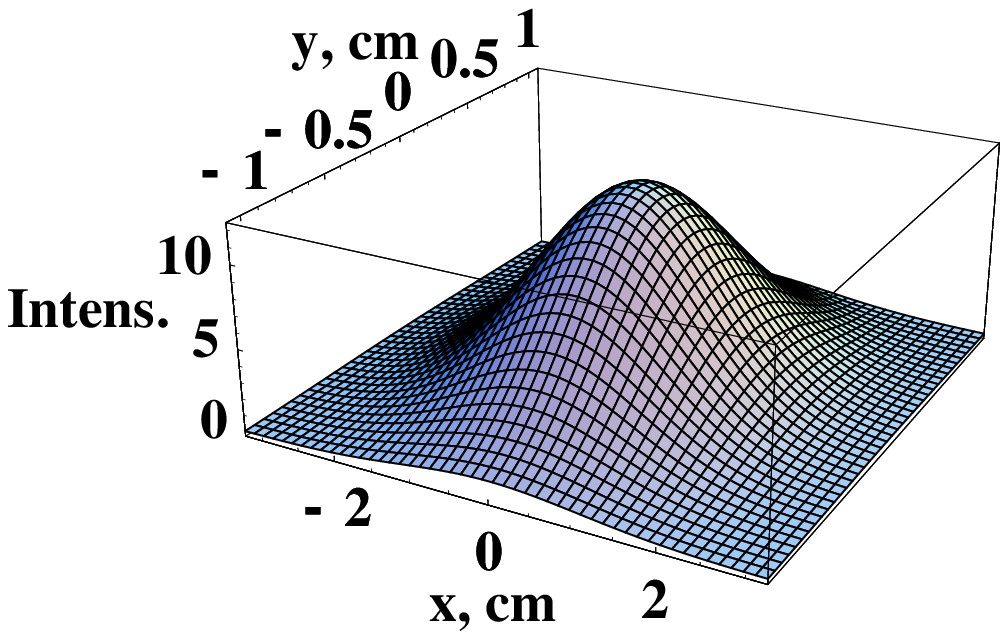}
 \hfill
\includegraphics[width=0.48\textwidth]
{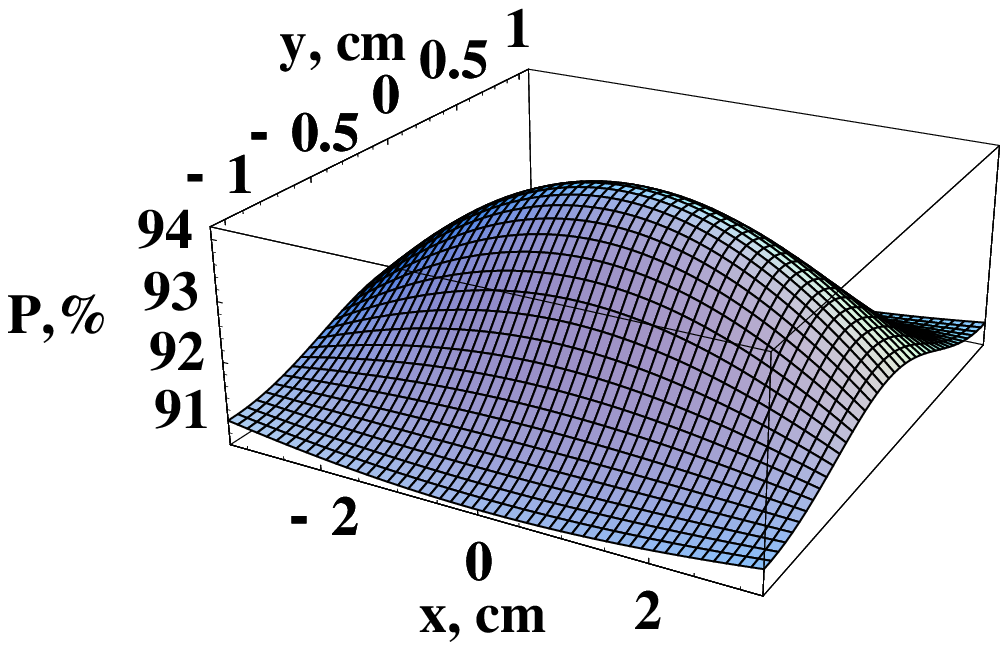}
 \hfill
 \parbox[t]{0.48\textwidth}{\caption {Spatial distribution of
 photon beam intensity. (Arbitrary units.) Photon energy
 220 MeV$<\omega_2<$330 MeV, $\xi^1_3 = +1$.
} \label{fig8}}
 \hfill
\parbox[t]{0.48\textwidth}{\caption{
 Degree of photons' linear polarization, $P$, \%. Photon energy
 220 MeV$<\omega_2<$330 MeV, $\xi^1_3 = +1$.
 }\label{fig9}}
\end{figure}
 \begin{figure}[ht]
\includegraphics[width=0.48\textwidth]
{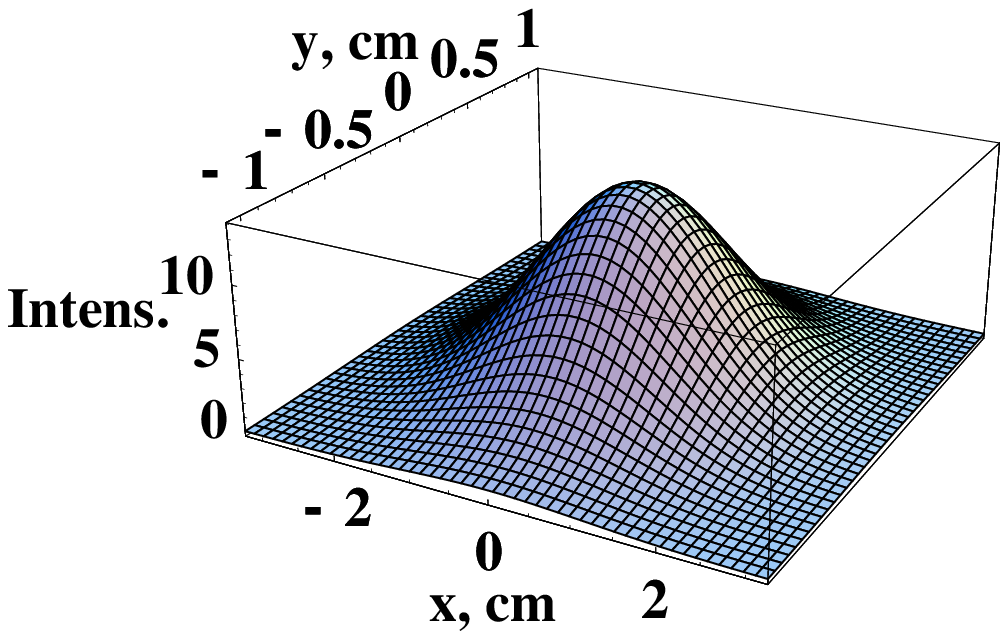}
 \hfill
\includegraphics[width=0.48\textwidth]
{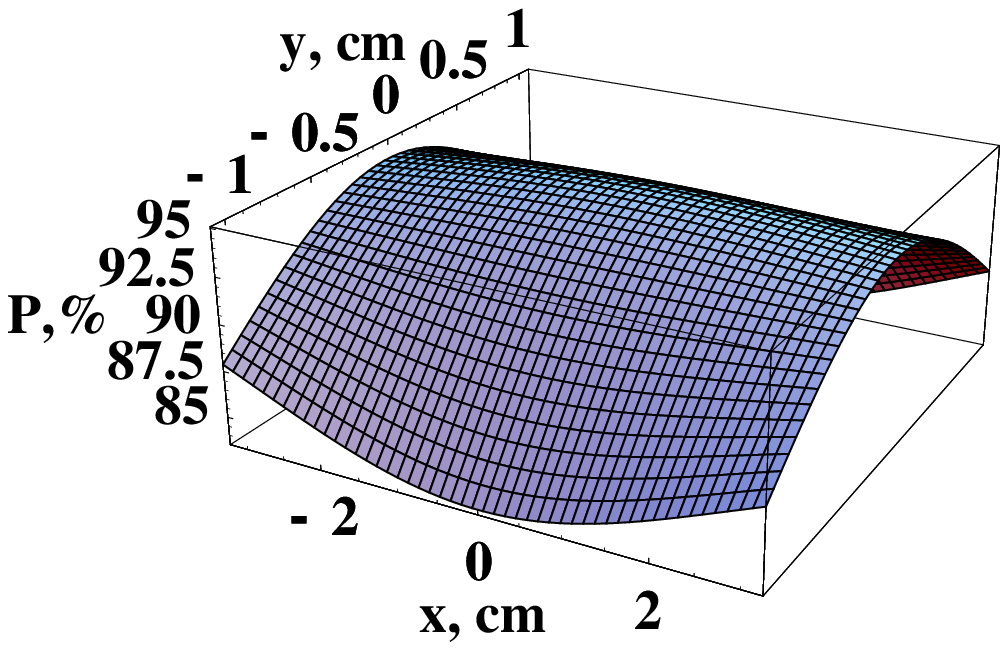}
 \hfill
 \parbox[t]{0.48\textwidth}{\caption {Spatial distribution of
 photon beam intensity. (Arbitrary units.) Photon energy
 220 MeV$<\omega_2<$330 MeV, $\xi^1_3 = -1$.
} \label{fig10}}
 \hfill
\parbox[t]{0.48\textwidth}{\caption{
 Degree of photons' linear polarization, $P$, \%. Photon energy
 220 MeV$<\omega_2<$330 MeV, $\xi^1_3 = -1$.
 }\label{fig11}}
\end{figure}

 \begin{figure}[t]
\includegraphics[width=0.48\textwidth]
{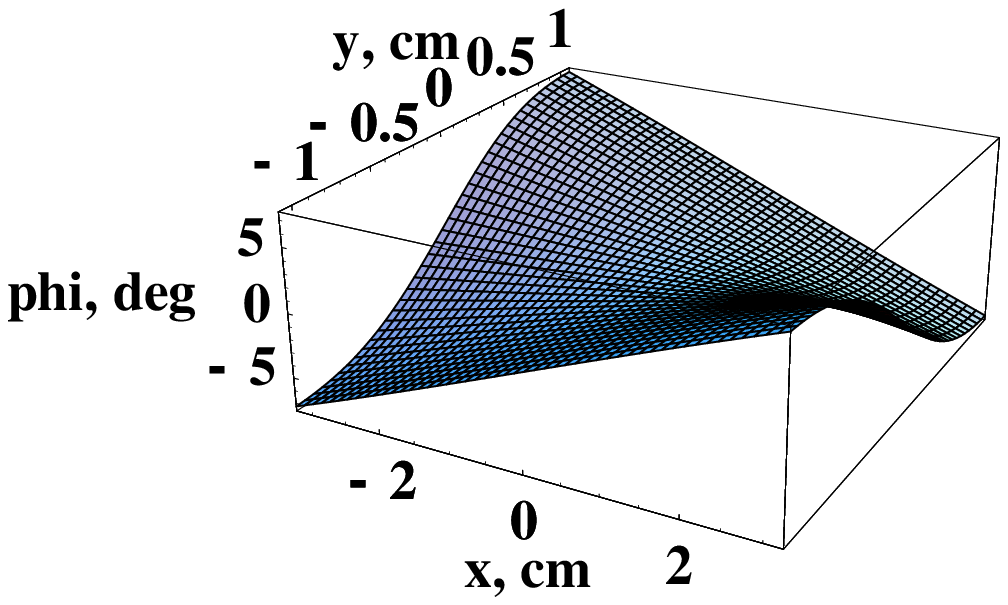}
 \hfill
\includegraphics[width=0.48\textwidth]
{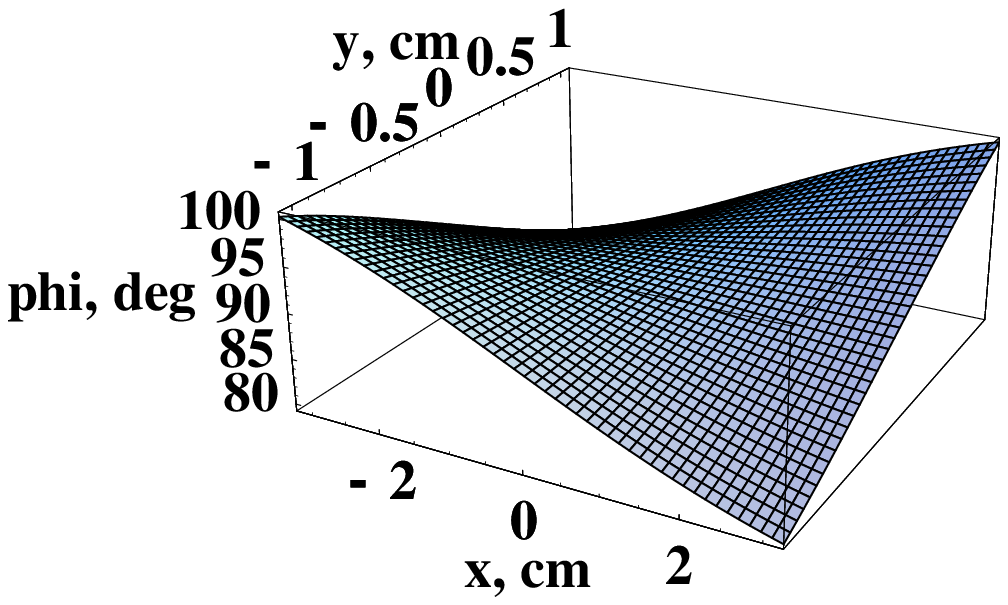}
 \hfill
 \parbox[t]{0.48\textwidth}{\caption {Spatial distribution of
 angle $\phi(x.y)$ between axis $x$ and direction of photon
 linear polarization Photon energy
 220 MeV$<\omega_2<$330 MeV, $\xi^1_3 = +1$.
} \label{fig10}}
 \hfill
\parbox[t]{0.48\textwidth}{\caption{
Spatial distribution of
 angle $\phi(x.y)$ between axis $x$ and direction of photon
 linear polarization Photon energy
 220 MeV$<\omega_2<$330 MeV, $\xi^1_3 = -1$.
 }\label{fig11}}
\end{figure}

\newpage
%\clearpage

\end{document}